# Discovery of Intrinsic Ferromagnetic Ferroelectricity in Transition Metal Halides Monolayer

Chengxi Huang[1], Yongping Du[1], Haiping Wu[1,*], Hongjun Xiang [3,4,*], Kaiming Deng[1], Erjun Kan[1,2,*]

[1] *Department of Applied Physics and Institution of Energy and Microstructure, Nanjing University of Science and Technology, Nanjing, Jiangsu 210094, P. R. China*

[2] *Key Laboratory of Soft Chemistry and Functional Materials (Ministry of Education), Nanjing University of Science and Technology, Nanjing, Jiangsu 210094, P. R. China*

[3] *Key Laboratory of Computational Physical Sciences (Ministry of Education), State Key Laboratory of Surface Physics, and Department of Physics, Fudan University, Shanghai 200433, P. R. China*

[4] *Collaborative Innovation Center of Advanced Microstructures, Nanjing 210093, P. R. China*

★Correspondence and requests for materials should be addressed to

E. K. (email: ekan@njust.edu.cn), H. X. (email: hxiang@fudan.edu.cn), H. W. (email: mrhpwu@njust.edu.cn)

The realization of multiferroics in nanostructures, combined with a large electric dipole and ferromagnetic ordering, could lead to new applications, such as high-density multi-state data storage. Although multiferroics have been broadly studied for decades, ferromagnetic ferroelectricity is rarely explored, especially in two-dimensional (2D) systems. Here we report the discovery of 2D ferromagnetic ferroelectricity in layered transition metal halide systems. On the basis of first-principles calculations, we reveal that charged $CrBr_3$ monolayer exhibits in-plane multiferroicity, which is ensured by the combination of orbital and charge ordering as realized by the asymmetric Jahn-Teller distortions of octahedral Cr-$Br_6$ units. As an example, we further show that $(CrBr_3)_2Li$ is a ferromagnetic ferroelectric multiferroic. The explored phenomena and mechanism of multiferroics in this 2D system are not only useful for fundamental research in multiferroics but also enable a wide range of applications in nano-devices.

Multiferroics, which simultaneously exhibit ferroelectricity and magnetism, have received intensive studies because of their novel physics and potential applications for spintronics and memory devices [1-4]. Until now, two kinds of multiferroics have been well discovered. In the type-I multiferroics, the ferroelectricity and magnetic ordering have different sources, which usually results in a weak magnetoelectric (ME) coupling. In the type-II multiferroics, the ferroelectricity is caused by the magnetic ordering which breaks the centrosymmetry. In this case, the ME coupling is strong, but the electric polarization is usually very small and the transition temperature is quite low [3,4]. Two-dimensional (2D) materials with robust ferroelectric (FE) as well as ferromagnetic (FM) order are promising materials for multi-functional applications but are rather scarce [5,6] because of the inherent exclusion between ferroelectricity and ferromagnetism [3,7]. Actually, up to now, 2D FM FE systems have not been reported yet.

Recently, ferroelectricity in 2D materials was discovered. For example, 1*T* $MoS_2$ [8], functionalized graphene [9], SnSe [10], phosphorene [11,12], phosphorus oxides [13], and $In_2Se_3$ [14] are theoretically predicted to be 2D FE materials. Moreover, Chang *et al.* have experimentally shown that tin telluride (SnTe) monolayer has robust FE properties [15]. Also, 2D ferromagnetism has been experimentally observed in two transition metal halide (TMH) layered systems, i.e. $Cr_2Ge_2Te_6$ [16] and $CrI_3$ [17]. The family of TMHs has received much attention because of their interesting topological properties [17-25]. Note that most of the TMHs are magnetic semiconductors or insulators. These exciting discoveries have motivated us to explore the possibility of coexistence of ferromagnetism and ferroelectricity in a single 2D system, e.g., a TMH materials

Here, in this work, by studying an example system of the TMH monolayers ($CrBr_3$), we discover that multiferroicity could be induced by the combination of charge-order (CO) and orbital-order (OO). Based on first-principles calculations, our results show that, by doping one electron in the $CrBr_3$ primitive cell, the anomalous asymmetric Jahn-Teller (JT) distortions of two neighboring Cr-$Br_6$ units bring simultaneously CO and OO. The resulting spatial electron-hole separation and spontaneous symmetry breaking will lead to 2D multiferroicity in the $CrBr_3^{0.5-}$ system. We further investigate the FM and FE properties of $(CrBr_3)_2Li$, in which the $Li^+$ cations are adopted as electron donors. Finally, we discuss the possible multiferroicity in other TMHs systems.

Our first-principles calculations are based on density functional theory (DFT) as implemented in the Vienna *Ab initio* Simulation Package (VASP) [26]. Exchange-correlation functional proposed by Perdew, Burke, and Ernzerh (PBE) [27] was used. The effective Hubbard $U$ = 3 eV was added according to Dudarev's method [28] for the Cr-*d* orbitals. The projector augmented wave (PAW) method [29] was used to treat the core electrons. The plane-wave cutoff energy was set to be 500 eV and the first Brillouin zone was sampled using a Γ-centered 12×12×1 Monkhorst-Pack grid [30]. A vacuum space of 20 Å was adopted and the FE polarization was calculated using the Berry phase method [31].

Ferroelectricity originates from the separation between negative and positive charges. Previous studies have shown that either CO or OO [32-36] can lead to ferroelectricity in 3D systems. However, in some systems, individual CO or OO cannot ensure the appearance of ferroelectricity. For instance, in a 2D honeycomb lattice with a uniform charge distribution (Fig. 1), the existence of centrosymmetry makes it non-polar. If $\delta$ charges transfer from one sublattice to the other one, a CO phase is formed, but the system is still non-polar because of the in-plane $c_{3\perp}$ symmetry. Then if we further induce OO in this system, which breaks the in-plane $c_{3\perp}$ symmetry, an in-plane electric polarization will occur. Thus, in this case, the combination of CO and OO is responsible for the appearance of ferroelectricity.

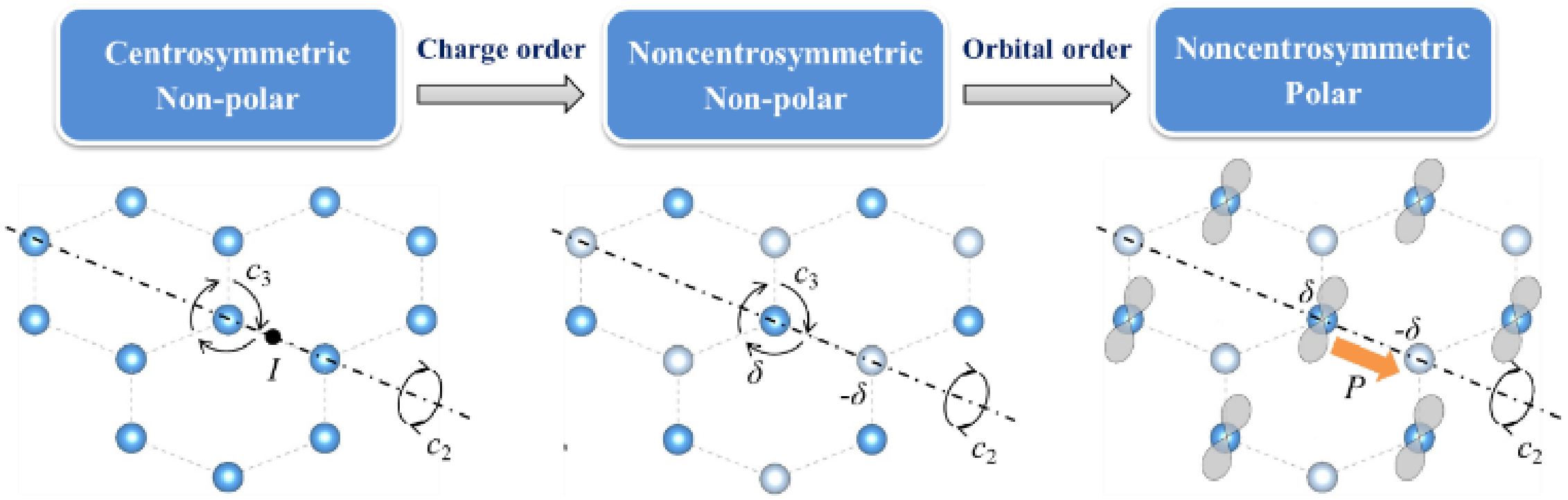

**FIG. 1.** Evolution from a non-polar to a polar honeycomb structure driven by the combination of charge-order (CO) and orbital-order (OO).

One of the candidate systems to study the CO-OO-cooperating ferroelectricity is $CrBr_3$. The $CrBr_3$ monolayer shown in Fig. 2(a) (point group $D_{3d}$) is a non-polar FM insulator. In the cases of hole doping, the $CrBr_3^{x+}$ ($0 < x \leq 0.5$) systems keep the $D_{3d}$ symmetry. While for electron doping, a remarkable JT distortion in $CrBr_3^{x-}$ appears, resulting in a structure with a polar $C_2$ symmetry. This dramatically difference can be understood. In the case of hole doping, the Fermi level of $CrBr_3$ shifts downward into the bands dominated by Br-*p* orbitals. Thus no JT distortions appear. While in the case of electron doping, the Fermi level shifts upward, leading to partially occupied Cr-*e* states, where JT distortions occur to break the degeneracy and stabilize the system [Fig. 2(f)].

The optimal structure of $CrBr_3^{0.5-}$ is shown in Fig. 2(b). Interestingly, the distortions of two neighboring Cr-$Br_6$ units are not equal. In the Cr1-$Br_6$ unit, the two Cr-Br bonds along z-direction (~ 3.0 Å) are much longer than the other four (~ 2.6 Å). While in the Cr2-$Br_6$ unit, the distortions are less obvious. This is quite anomalous compared to the common JT distortions in pristine TMHs [37]. We name these two kinds of distortions "asymmetric distortions" and "symmetric distortions", respectively. We also calculated the $CrBr_3^{0.5-}$ systems with fixed $D_{3d}$ (with no JT distortions) and $C_{2h}$ (with symmetric JT distortions) symmetries. The results show that both $D_{3d}$- (0.26 eV) and $C_{2h}$-$CrBr_3^{0.5-}$ (0.11 eV) are much higher in energy than the $C_2$-$CrBr_3^{0.5-}$ (Fig. S1 [38]). Moreover, our DFT +$U$ ($U$ = 1 to 4 eV) and HSE06 calculations give similar results, and the dynamical stability of $C_2$-$CrBr_3^{0.5-}$ is verified by phonon calculations (Fig. S2 [38]).

The formal oxidation state of Cr1 and Cr2 is +2 and +3, with magnetic moments of 4 and 3 $\mu_B$, respectively. An CO phase is clearly observed [Fig. 2(c)]. The FM order is more favored than the antiferromagnetic (AFM) order (two neighboring spins are antiparallel). Zhou *et al.* has demonstrated that electron doping could enhance the FM coupling in a 2D organometallic system [47]. While in $CrBr_3$, electron doping causes asymmetric JT distortions, greatly changing the electronic structures (e.g. split of Cr-e states) near the Fermi level as well as the magnetic coupling.

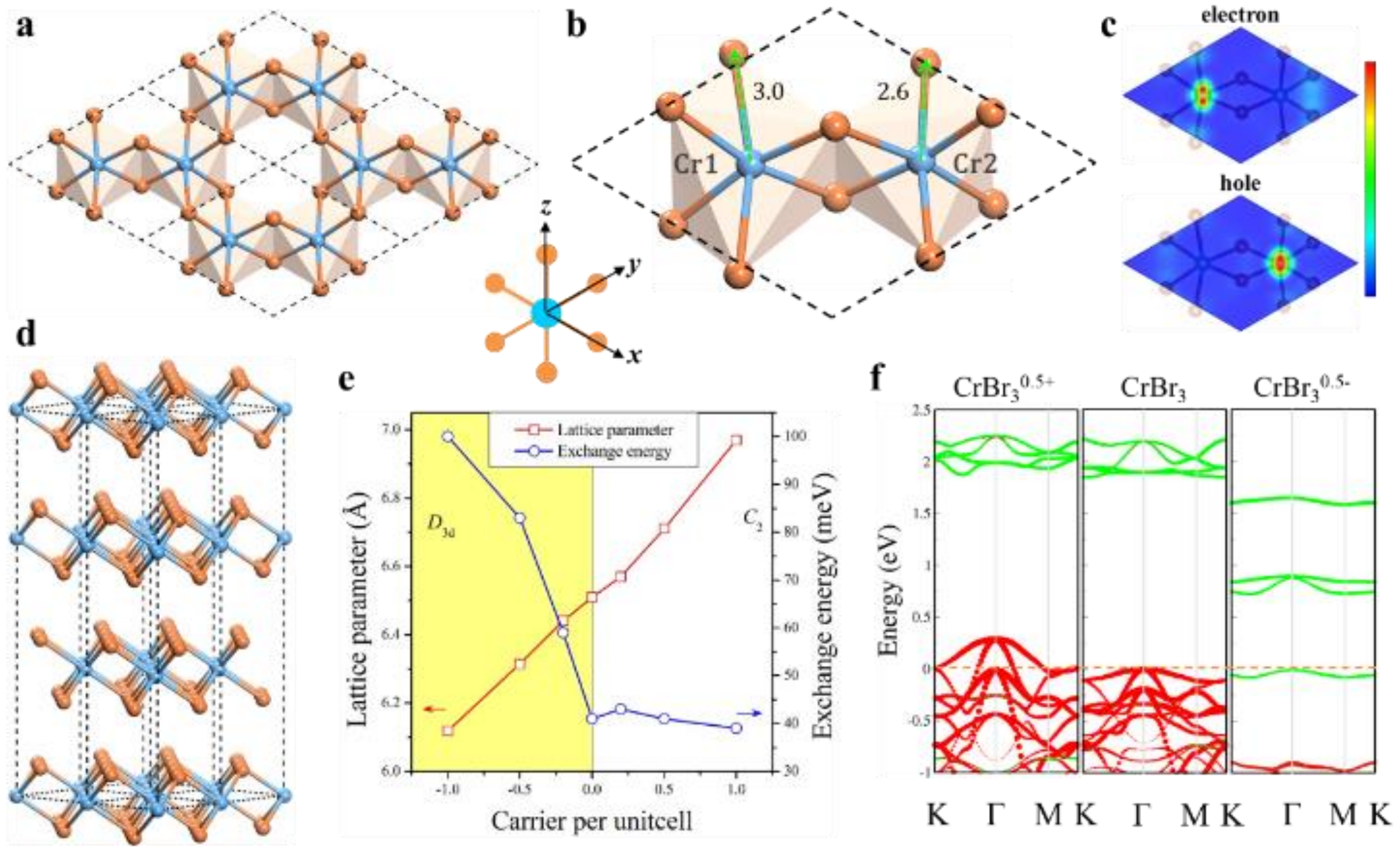

**FIG. 2.** Top view of the optimized $CrBr_3$ (a) and $CrBr_3^{0.5-}$ (b) monolayer. Brown and cyan balls represent Br and Cr atoms, respectively. Dashed rhombus represents the primitive cell. (c) Electron and hole density in real space for the valence band maximum and the conduction band minimum, respectively. Red and blue represent

the maximum and minimum value, respectively. (d) Pristine $CrBr_3$ bulk. (e) Lattice parameter and exchange energy ($E_{AFM}$ – $E_{FM}$) per unit cell refer to carrier concentration. Negative and positive value of carrier concentration represents hole and electron doping, respectively. (f) Orbital-resolved band structures for $CrBr_3^{0.5+}$, $CrBr_3$ and $CrBr_3^{0.5-}$. Green lines represent Cr-$d$ states. Red lines represent Br-$p$ states.

Due to the asymmetric JT distortions, the electric polarization of $C_2$-$CrBr_3^{0.5-}$ occurs parallel to the in-plane $c_2$ axis, which is perpendicular to the occupied Cr1-$dz^2$ orbital. To reveal the origin of the in-plane polarization, we considered two similar hypothetical systems, i.e., $CrFeBr_6$ (replace one Cr to Fe atom in the $CrBr_3$ unit cell) and $MnBr_3$ (replace all Cr to Mn atoms). The optimized structure of $CrFeBr_6$ belongs to $D_3$ symmetry, where no JT distortion is observed. The magnetic moments of $Fe^{3+}$ and $Cr^{3+}$ are 5 and 3 $\mu_B$, respectively. Thus the $D_3$-$CrFeBr_6$ can be considered to have CO but no OO. For $MnBr_3$, the optimized structure belongs to $C_{2h}$ symmetry and JT distortions equally appear in each Mn-$Br_6$ unit. The magnetic moment of each $Mn^{3+}$ is 4 $\mu_B$. Thus the $C_{2h}$-$MnBr_3$ has OO but no CO. While in $C_2$-$CrBr_3^{0.5-}$, because of the asymmetric JT distortions, CO and OO phases are simultaneously formed. However, we find that both $D_3$-$CrFeBr_6$ and $C_{2h}$-$MnBr_3$ are non-polar, only $C_2$-$CrBr_3^{0.5-}$ is polar. Therefore, the in-plane polarization of $C_2$-$CrBr_3^{0.5-}$ is driven by the combination of CO and OO. To verify these spontaneous symmetry breaking in $CrBr_3^{0.5-}$, we also plotted the phonon dispersions of the non-polar phases, namely $D_{3d}$-, $C_{2h}$- and $D_3$-$CrBr_3^{0.5-}$ (Fig. S2 [38]), in which the imaginary bands of soft optical modes are observed [48]. The relation between JT distortions and ferroelectricity has also been studied in perovskite systems [49], which may allow electric field-control of the structural and electronic properties.

To explain why the asymmetric JT distortions are preferred over the symmetric JT distortions in $CrBr_3^{0.5-}$, we start from the degenerated Cr-$e$ states derived from the octahedral ligand field. In $C_{2h}$-$CrBr_3^{0.5-}$, Cr-*e* states of two Cr-$Br_6$ units equally split to higher $dx^2$-$y^2$ and lower $dz^2$ states. The added electron occupies the bonding state generated from the hybridization between Cr1- and Cr2-$dz^2$ states [Fig. 3(a) and (c)]. The average valence for Cr is +2.5. In this case, the electronic energy benefits from two parts, i.e. split of Cr-*e* states and hybridizations between Cr-$dz^2$ orbitals. In $C_2$-$CrBr_3^{0.5-}$, distortions in the Cr1-$Br_6$ unit are much stronger than that in Cr2-$Br_6$, leading to unequal split of two Cr-*e* states. The lower Cr1-$dz^2$ state hardly hybridize with the higher Cr2-$dz^2$ state. Then the Cr1-$dz^2$ state is occupied by the added electron, while the Cr2-$dz^2$ state is empty [Fig. 3(b) and (d)]. The electronic energy only benefits from the split of Cr1-*e* state. As we know that the 3*d* orbitals are much localized. Thus, the direct hybridizations between Cr-$dz^2$ orbitals can be omitted. The electronic energy benefits is mainly decided by the strength of split of Cr-*e* states. From Fig. 3(c) and (d), one finds that the split of Cr1-*e* states from asymmetric distortions (~ 1.6 eV in energy) is much larger than that from symmetric distortions (~ 0.8 eV in energy), which explains why asymmetric distortions are preferred in this system.

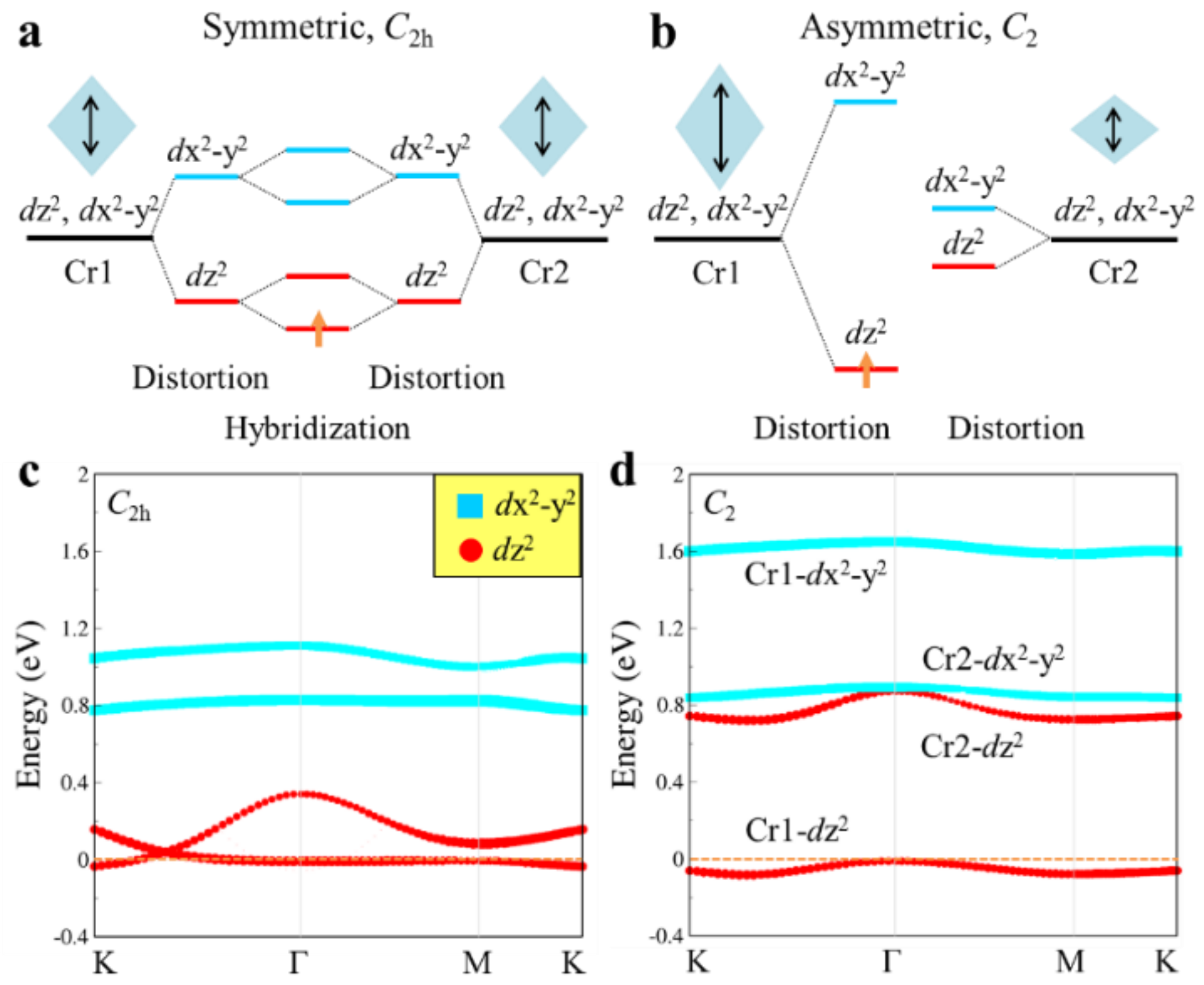

**FIG. 3.** Evolution of states under symmetric (a) and asymmetric (b) JT distortions. Blue rhombuses represent the distorted octahedral ligand field for each Cr-$Br_6$ sublattice. Orbital-resolved bands around the Fermi level for $CrBr_3^{0.5-}$ with $C_{2h}$ (c) and $C_2$ (d) symmetry.

Magnetic anisotropy (MA) and ME coupling in $C_2$-$CrBr_3^{0.5-}$ is examined by including the spin-orbit coupling (SOC). Pristine $CrBr_3$ monolayer has an out-of-plane magnetic easy axis [24]. However, in $C_2$-$CrBr_3^{0.5-}$, the easy axis is within in the atomic plane, parallel to the in-plane polarization. This implies that the in-plane FE order is coupled with the net magnetization. Moreover, according to the combinations of different CO and OO, there should be six possible orientations of the in-plane FE order, i.e. 0°, 60°, 120°, 180°, 240° and 300° orientations (Fig. S3 [38]).

After studying of the hypothetical $CrBr_3^{0.5-}$ system, we now turn to the realistic materials feasible in experiments. There are many ways to doping electrons into a 2D system, such as using a substrate, doping metals or ions. Lei *et al.* have achieved a high-density (Li/Fe ratio up to 0.5) carrier doping in Layered FeSe thin film by using $Li^+$ [50]. Here we also choose $Li^+$ as electron donors. Four possible adsorption sites of $Li^+$ on $CrBr_3$ monolayer are considered (Fig. S4 [38]), and the hole center of the porous $CrBr_3$ framework is preferred by the dopant [Fig. 4(a)]. In $(CrBr_3)_2Li$ ($C_2$ point group), large asymmetric distortions take place in the $CrBr_3$ framework. The lateral lattice parameter is 6.49 Å and the angle between *a* and *b* lattice vectors is 118.6°. Its thermal stability is examined by performing *ab* initio molecular dynamics simulations at 300 K (Fig. S6 [38]).

To further confirm the ferroelectricity in $(CrBr_3)_2Li$, we considered four possible configurations [namely, FE, antiferroelectric (AFE) and ferrielectric (FIE1 and FIE2) orders, see Fig. 4(b)] in a 2 × 1 supercell containing four $Cr\text{-}Br_6$ units. The AFE structure is found to have higher energy (~21.3 meV per Cr) than the FE structure, while the FIE1 and FIE2 structures always automatically converge into the FE and AFE structures, respectively. Furthermore, we considered two typical orders in a $\sqrt{3} \times \sqrt{3}$ supercell containing six $Cr\text{-}Br_6$ units, namely, the 120° AFE and FIE3 order [Fig. 4(b)]. The relative energies of 120° AFE and FIE3 structure are ~ 9.0 and ~ 22.3 meV per Cr with respect to the FE structure. We have also considered two other similar $CrBr_3^{0.5-}$ systems, namely $(CrBr_3)_2NH_4$ and $(CrBr_3)_2Na$ (Fig. S4, S5, S6 and S7 [38]), and find that the 120° AFE structures have the lowest energy (Table S2 [38]). These imply that whether the FE order is favored in $CrBr_3^{0.5-}$ systems or not depends on the properties (e.g., size) of the electron donors.

Same as $CrBr_3^{0.5-}$, the magnetic ground state of FE $(CrBr_3)_2Li$ monolayer is FM with exchange energy ($\Delta E = E_{\mathrm{AFM}} - E_{\mathrm{FM}}$) of ~ 15 meV per Cr. The magnetic moments of two neighboring Cr are 4 and 3 $\mu_{\mathrm{B}}$, respectively. Including SOC, the easy axis is parallel to the in-plane polarization. The MA energies are 87 (with respect to in-plane hard-axis) and 82 μeV (with respect to out-of-plane hard-axis) per Cr, respectively. These values are larger than those of traditional metals such as Fe (1.4 μeV). Due to the non-negligible uniaxial MA, here we use the Heisenberg model including single ion anisotropy to describe the magnetic behavior of this system. The spin Hamiltonian

$$H = -J\sum_{\langle ij\rangle} S_i S_j - D\sum_{\langle i\rangle} S_{i,x0}^2$$

where $J$ represents the nearest-neighbor exchange coupling, $D$ represents the single-ion anisotropy parameter, $S_{i,x0}$ represents the component of $S_i$ along the in-plane easy-axis (x0) direction. The values of $J$ and $D$ are calculated to be ~ 1661 and ~ 14 μeV, respectively. The $T_C$ is estimated to be ~ 50 K by classic Monte Carlo (MC) simulations [51]. The behaviors of M_x0 and M_tot are very close, indicating that the MA in this system is large enough to stabilize the orientation of net magnetization below the $T_C$.

The in-plane spontaneous polarization of FE order are calculated to be $0.92\times10^{-10}$ C/m for $(CrBr_3)_2Li$. If we take the thickness to be 6 Å, which is close to the interlayer space of bulk $CrBr_3$, the value will be ~ 15 $\mu C/cm^2$. This value is comparable to those of recently reported 2D FE SnSe, phosphorene [10-12] and $BaTiO_3$ bulk [44,52]. As mentioned above, there are six possible orientations of in-plane FE order in the $(CrBr_3)_2Li$. Here we explore the representative 120° FE switching by nudged-elastic-band (NEB) method (Fig. S9 [38]). The switching barrier is estimated to be ~15 meV/unit-cell, comparable to that of tetragonal bulk $BaTiO_3$ (~18 meV/unit-cell from our calculations). The ME coupling in $(CrBr_3)_2Li$ is also studied by the 120° FE switching, during which the orientation of magnetization is controlled by the electric field. The ME coupling coefficient is estimated to be in the same order as those in $Fe/BaTiO_3$ [45] and $BiFeO_3/CoFe_2O_4$ interfaces [46] (see Fig. S10 [38] for details).

For practical interests, it is necessary to examine whether the inherent properties of a 2D material could be affected by the substrate. Here we choose two non-magnetic non-polar insulators, namely the β-tridymite $SiO_2$ and α-$InBr_3$ slabs to support the $(CrBr_3)_2Li$ monolayer. After structural optimizations, the significant asymmetric JT distortions and in-plane ferroelectricity are observed in the $(CrBr_3)_2Li$ layer. The ground states are FM and the Cr-$dz^2$ and Cr-$dx^2$-$y^2$ bands are located within the large energy gap of the substrates. Thus, we expect that, the FM-FE multiferroic features of the $(CrBr_3)_2Li$ monolayer can persist on a suitable substrate (see [38] for details).

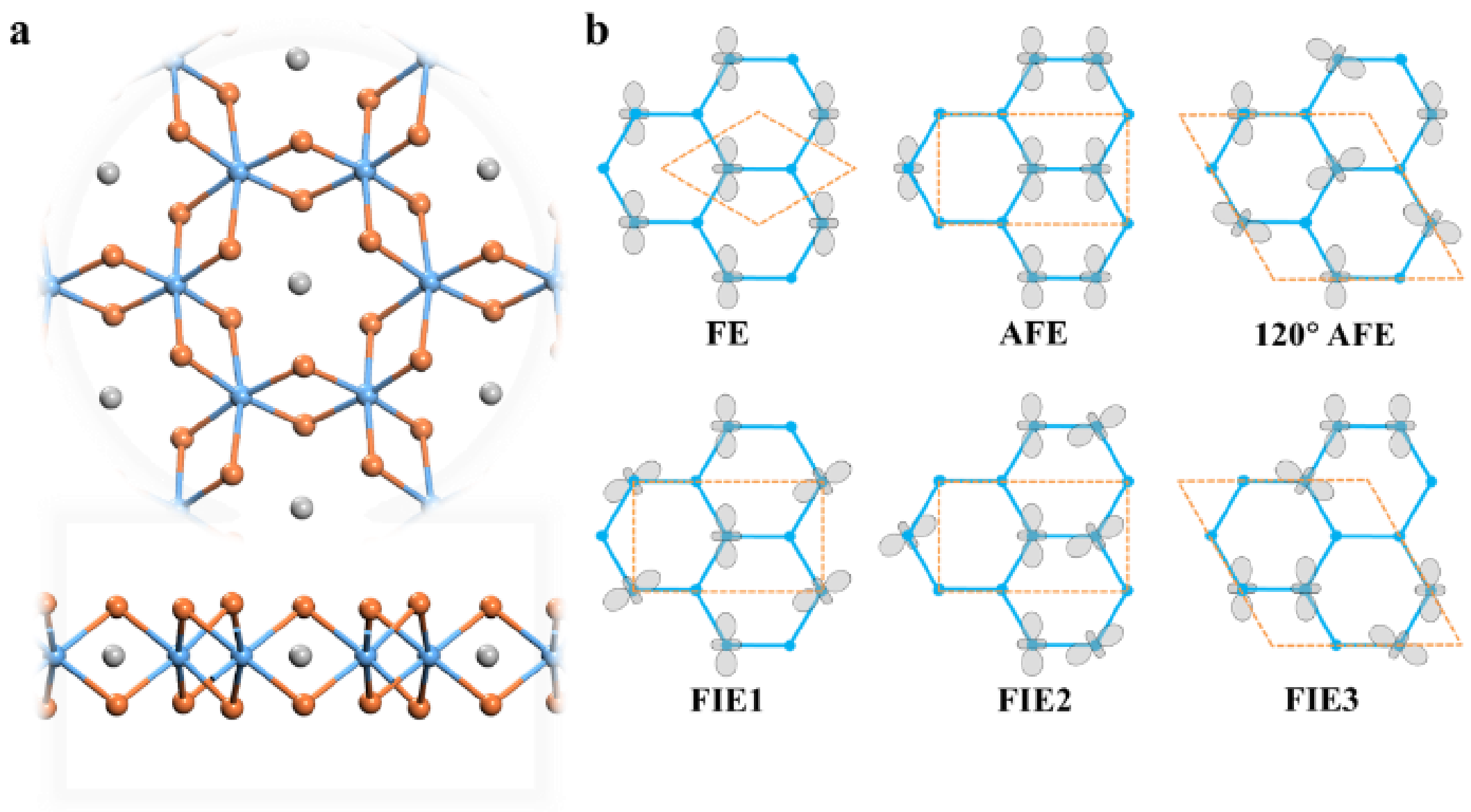

**FIG. 4.** (a) Top and side view of the optimized $(CrBr_3)_2Li$ monolayer. Brown, cyan and gray balls represent Br, Cr and Li atoms, respectively. (b) Schematic diagrams of ferroelectric (FE), antiferroelectric (AFE) and ferrielectric (FIE) orders. Blue framworks represent the honeycomb sublattice. Gray graphs represent the occupied Cr-$dz^2$ orbital. Orange dashed frames represent unit cells of structures with corresponding polarization orders.

At last, we discuss the possibility of multiferroicity in other charged TMH systems. First of all, $CrX_3$ (X = Cl, I) has very similar atomic (honeycomb framework consist of octahedral Cr-$X_6$ sublattices) and electronic structures (occupied $t_{2g}$ and empty $e_g$ in one spin channel) as $CrBr_3$. Our calculations show that the asymmetric JT distortions and multiferroic features also occur in $CrCl_3^{0.5-}$ and $CrI_3^{0.5-}$ (Fig. S12 [38]). Secondly, in some TMHs which exhibit inherent JT distortions with a mirror symmetry in their bulk phases such as $RhCl_3$ and $RhBr_3$, doping either electrons or holes could break the mirror symmetry and induce in-plane electric polarizations as well as magnetism. It appears that electronic correlations may play an important role in the occurrence of asymmetric instead of symmetric JT distortions. Overall, we suggest that 2D multiferroicity could widely exist in the family of TMHs, which may offer a new promising platform to study 2D multiferroics.

In summary, based on first-principles calculations, we reveal that the $CrBr_3^{0.5-}$ monolayer systems are 2D multiferroic semiconductors with in-plane FM and FE orders. The orientation of in-plane net magnetization is coupled with the electric polarization, evidencing the existence of ME coupling. An example material system (i.e. the $(CrBr_3)_2Li$ monolayer) is predicted to be a FM-FE multiferroic. In these systems, the in-plane ferroelectricity is ensured by the combination of CO and OO, which are induced by the asymmetric JT distortions of two neighboring Cr-$Br_6$ units. This mechanism is not restricted to the $CrBr_3^{0.5-}$ systems, but can also be applied to other TMHs and related 2D systems. These findings reveal the existence of multiferroicity in 2D systems and provide a new ideal platform to study 2D multiferroics. We are looking forward to future experimental realizations of 2D multiferroics in TMH systems.

E.K. is supported by the NSFC (51522206, 51790492,11574151, 11774173), by NSF of Jiangsu Province (BK20130031), by PAPD, the Fundamental Research Funds for the Central Universities (No.30915011203), and by New Century Excellent

Talents in University (NCET-12-0628). Work at Fudan is supported by NSFC, the Special Funds for Major State Basic Research (2015CB921700), Program for Professor of Special Appointment (Eastern Scholar), Qing Nian Ba Jian Program. C.H. and E.K. acknowledge the support from the Tianjing Supercomputer Centre and Shanghai Supercomputer Center. We thank Jian Zhou for valuable discussions.

# Supplementary Information

Discovery of Intrinsic Ferromagnetic Ferroelectricity in Transition Metal Halides Monolayer

## Section I: Properties of $CrBr_3^{0.5-}$.

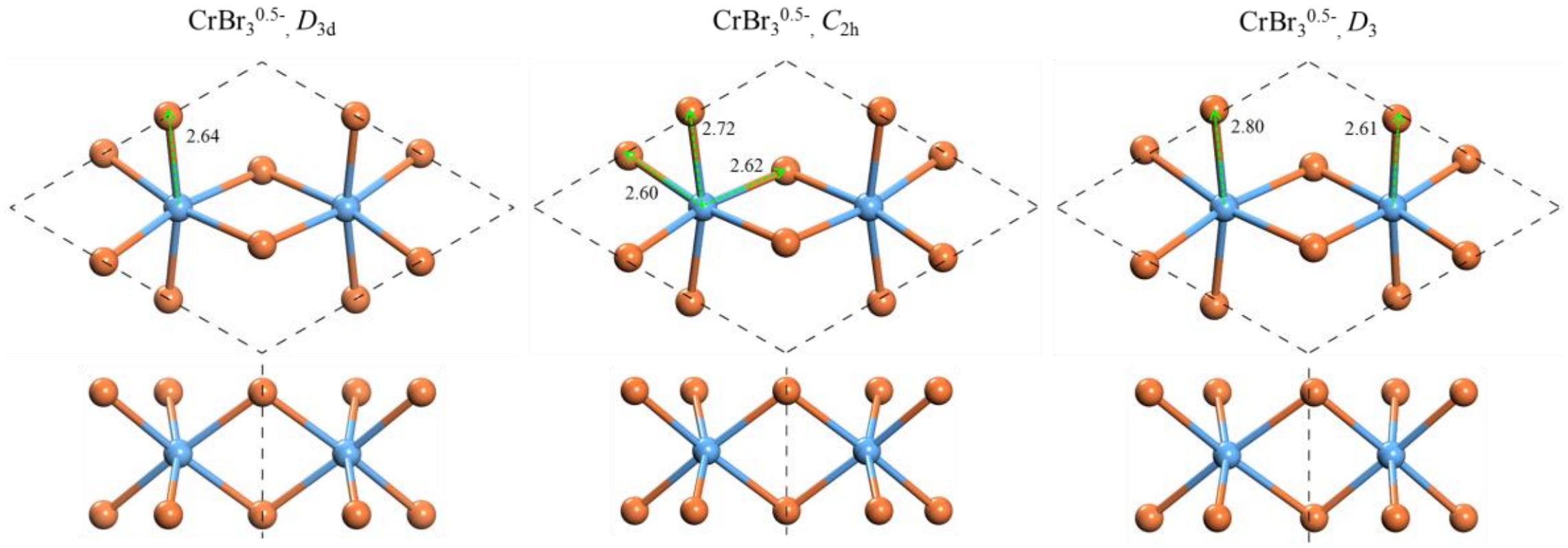

**Figure S1**. Structural details of $CrBr_3^{0.5-}$ with $D_{3d}$, $C_{2h}$ and $D_3$ symmetry. The Cr-Br bond lengths are in angstrom.

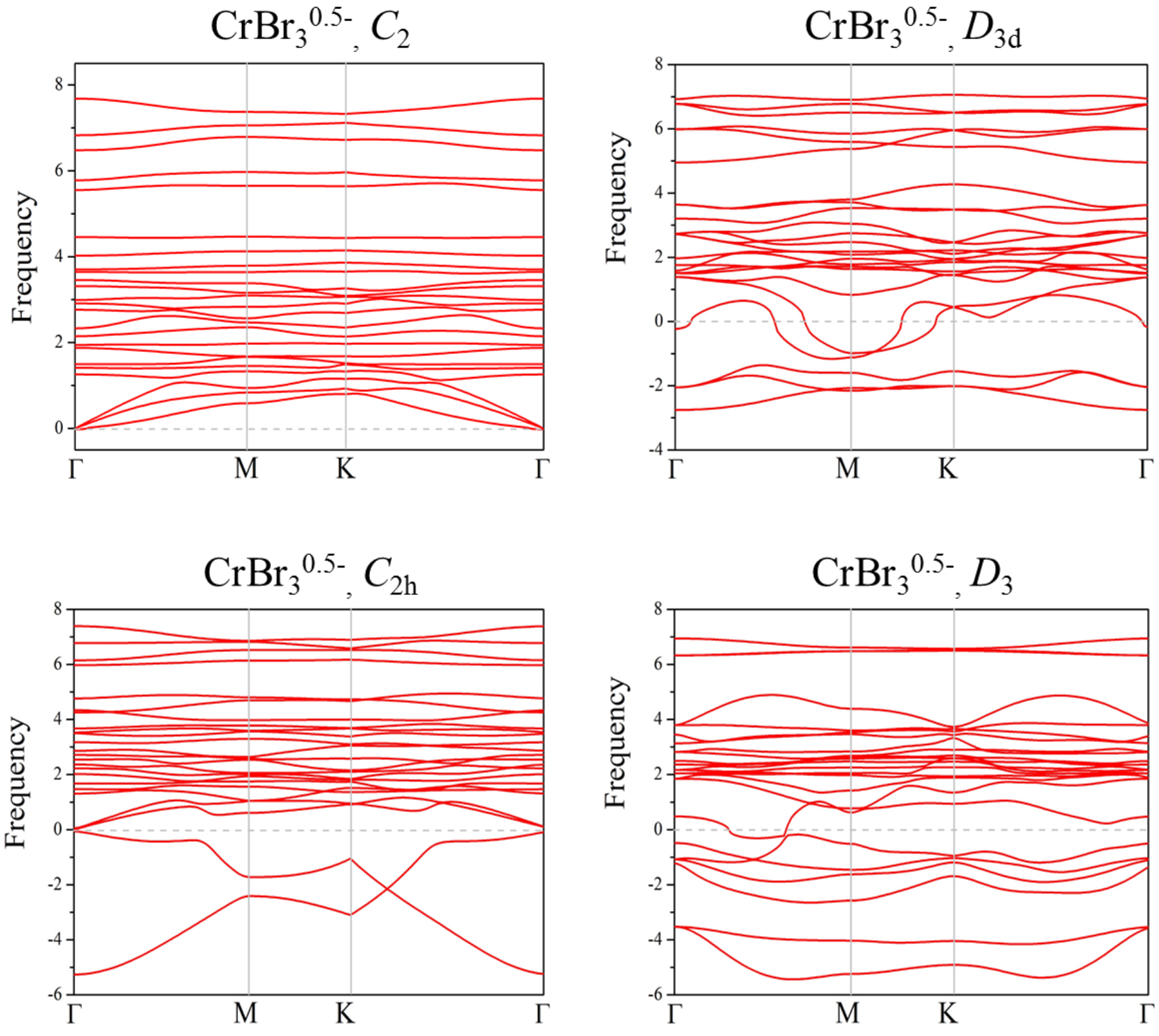

**Figure S2**. Phonon spectra for $C_2$-, $D_{3d}$-, $C_{2h}$- and $D_3$-$CrBr_3^{0.5-}$.

**Table S1**. The symmetry group of the optimal structure (*SG*), the energy difference between configurations with $C_{2h}$ and $C_2$ symmetry ($\Delta E_1$, defined as $\Delta E_1 = E(C_{2h}) - E(C_2)$) per unitcell, the energy difference between FM and AFM configuration ($\Delta E_2$, defined as $\Delta E_2 = E(\text{AFM}) - E(\text{FM})$) per Cr for $CrBr_3^{0.5-}$ under different value of Hubbard *U*.

| **Hubbard *U* (eV)** | 1 | 2 | 3 | 4 |
|---|---|---|---|---|
| ***SG*** | $C_2$ | $C_2$ | $C_2$ | $C_2$ |
| **$\Delta E_1$ (meV)** | 20 | 44 | 116 | 165 |
| **$\Delta E_2$ (meV)** | 24 | 22 | 19 | 17 |

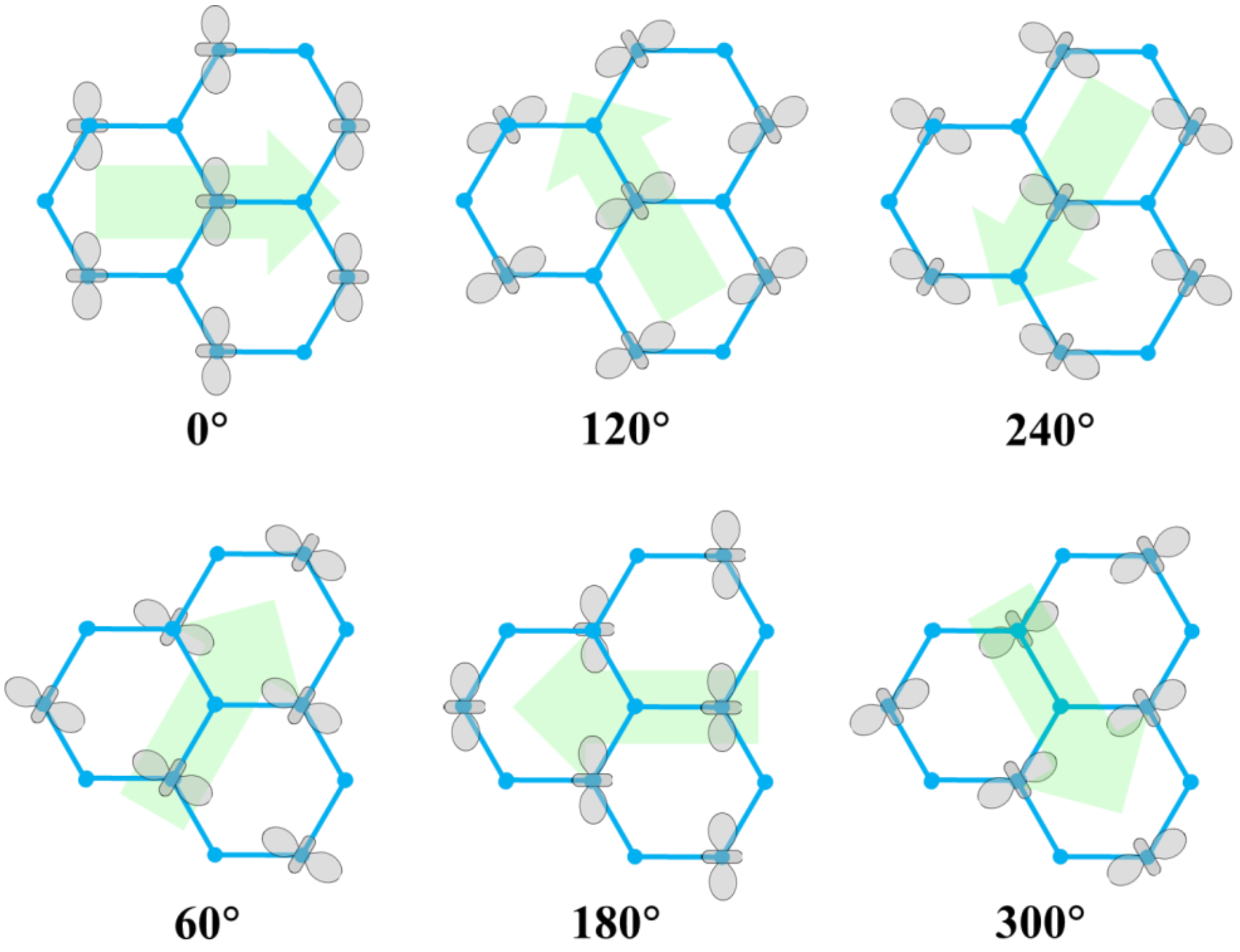

**Figure S3**. Six possible orientations of in-plane ferroelectric order. Blue framworks represent the honeycomb lattice of $CrBr_3^{0.5-}$. Gray graphs represent the occupied Cr-$dz^2$ orbital. Green arrows represent the in-plane electric polarizations.

## Section II: Stability, multiferroicity and electronic structures of $(CrBr_3)_2X$ (X = $NH_4$, Na, Li).

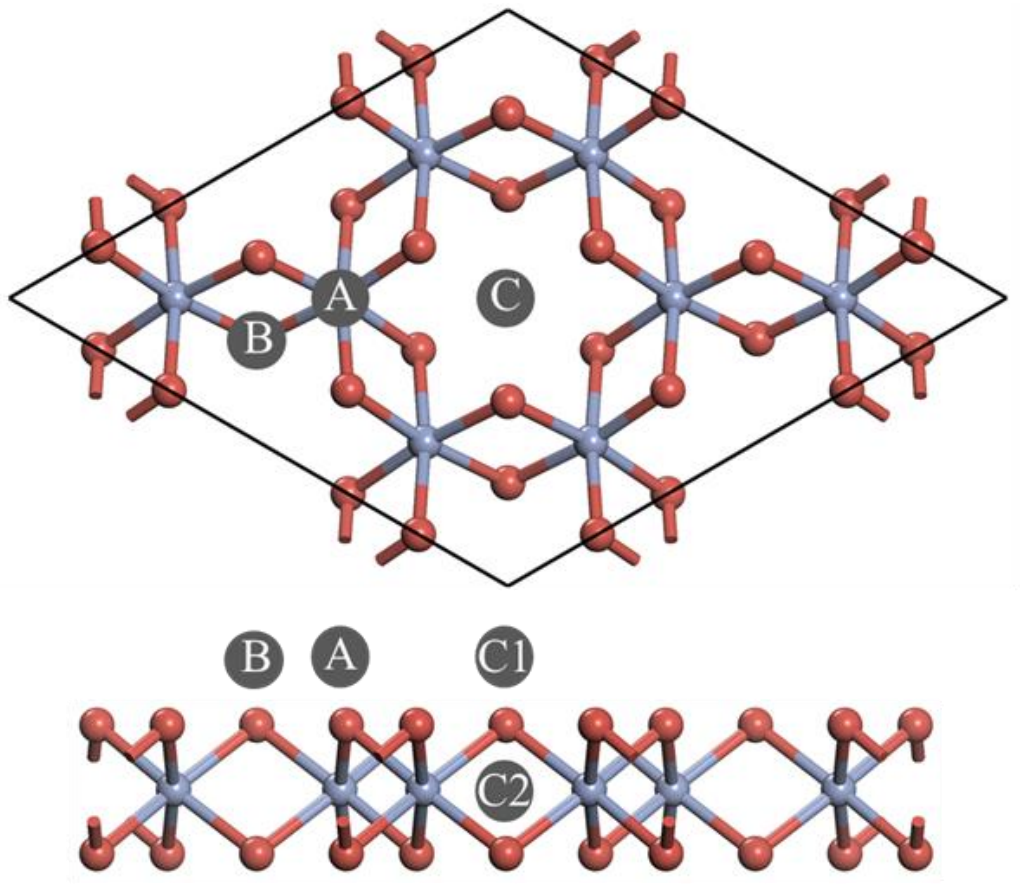

| Sites | Energy (eV) | | |
|---|---|---|---|
| | $NH_4^+$ | $Na^+$ | $Li^+$ |
| A(Cr-top) | 0.47 | 1.22 | 1.24 |
| B(Br-top) | 0.42 | 1.16 | 1.15 |
| C1(Hollow-top) | 0.27 | -- | -- |
| C2(Hollow) | 0 | 0 | 0 |

**Figure S4**. Four adsorption sites of $NH_4^+$, $Na^+$ and $Li^+$ on $CrBr_3$ monolayer and the corresponding energy per unitcell for each adsorption configuration. The total energy for C2 configuration is set to 0. The C1 intial structure of $(CrBr_3)_2Na$ and $(CrBr_3)_2Li$ always automatically converge into the C2 structure.

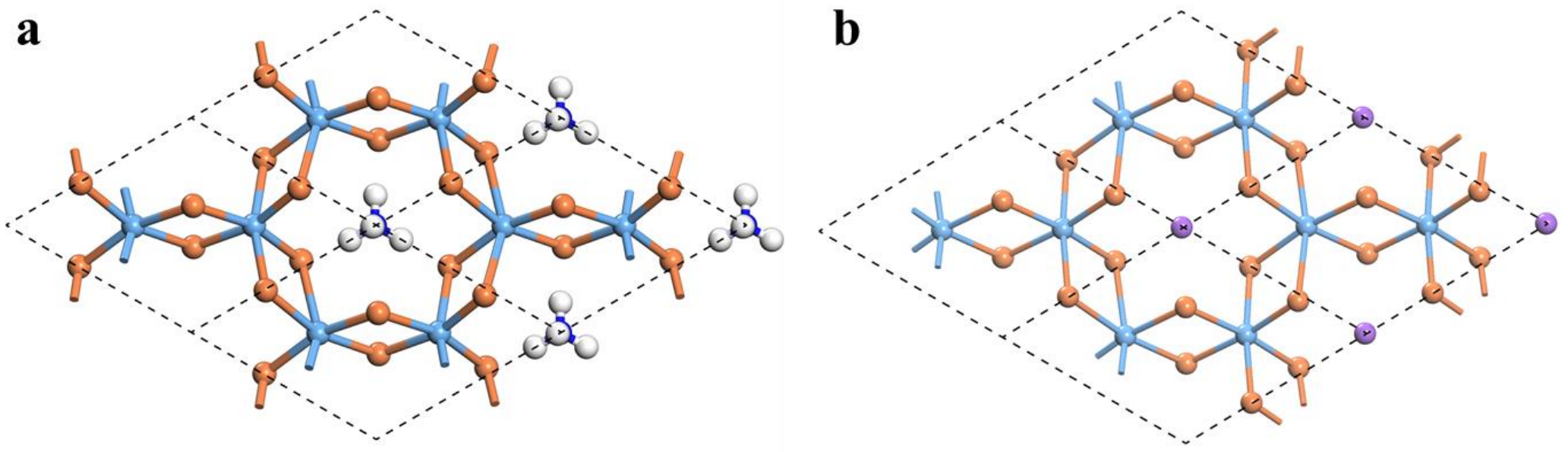

**Figure S5**. Atomic strucutres of (a) $(CrBr_3)_2NH_4$ and (b) $(CrBr_3)_2Na$. The 2D lattice parameters are 7.05 and 6.71 Å, and the angles between *a* and *b* lattice vectors are 119.5° and 118.8° for $(CrBr_3)_2NH_4$ and $(CrBr_3)_2Na$, respectively. $(CrBr_3)_2Na$ belongs to the $C_2$ point group.

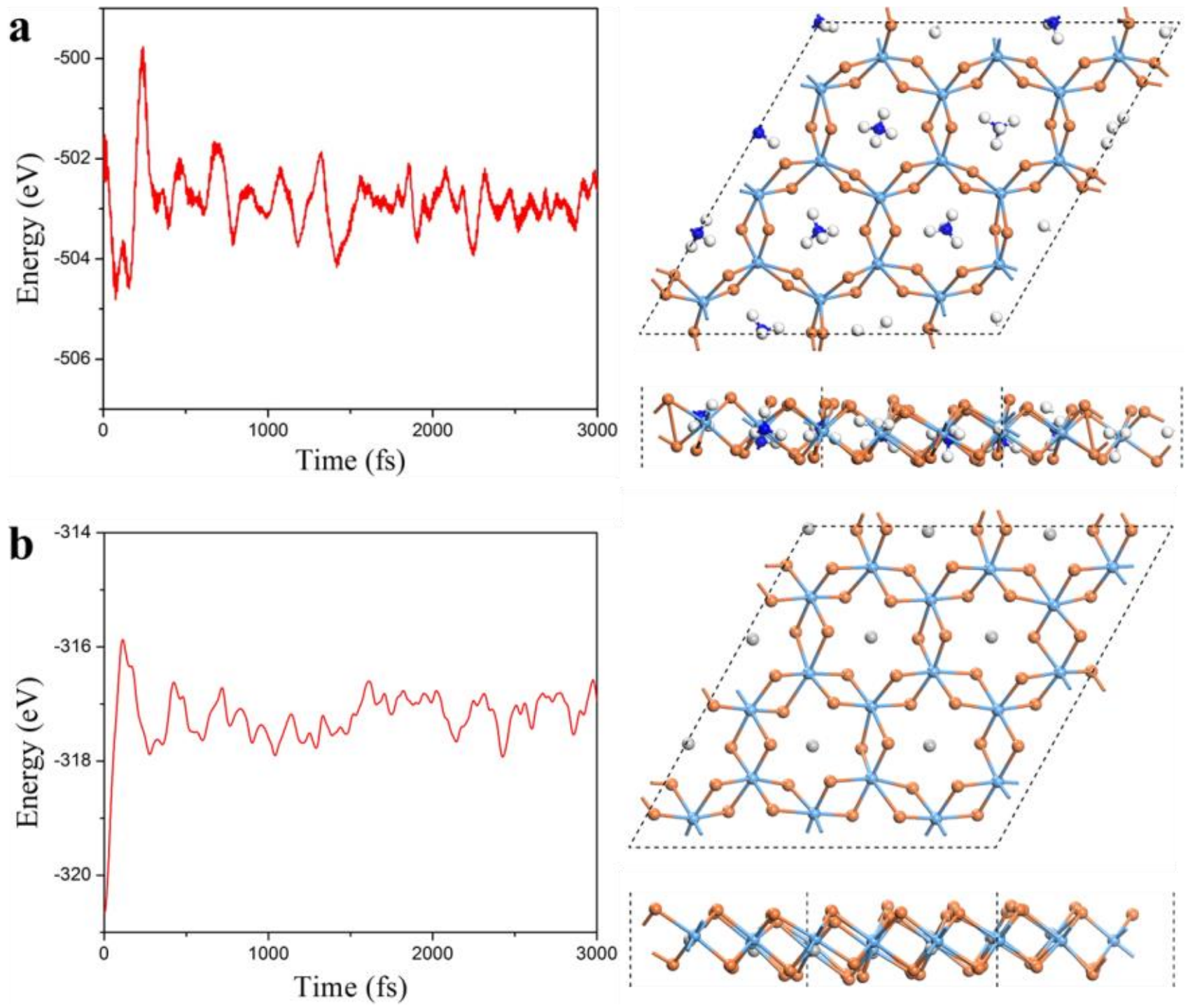

**Figure S6**. Total energy fluctuations with respect to molecular dynamics simulation step and structures after simulation time of 3000 fs for (a) $(CrBr_3)_2NH_4$ and (b) $(CrBr_3)_2Li$ monolayer. The *ab* initio molecular dynamics simulations (AIMD) are performed by using a 3×3×1 supercell under a Nose-Hoover thermostat at 300 K.

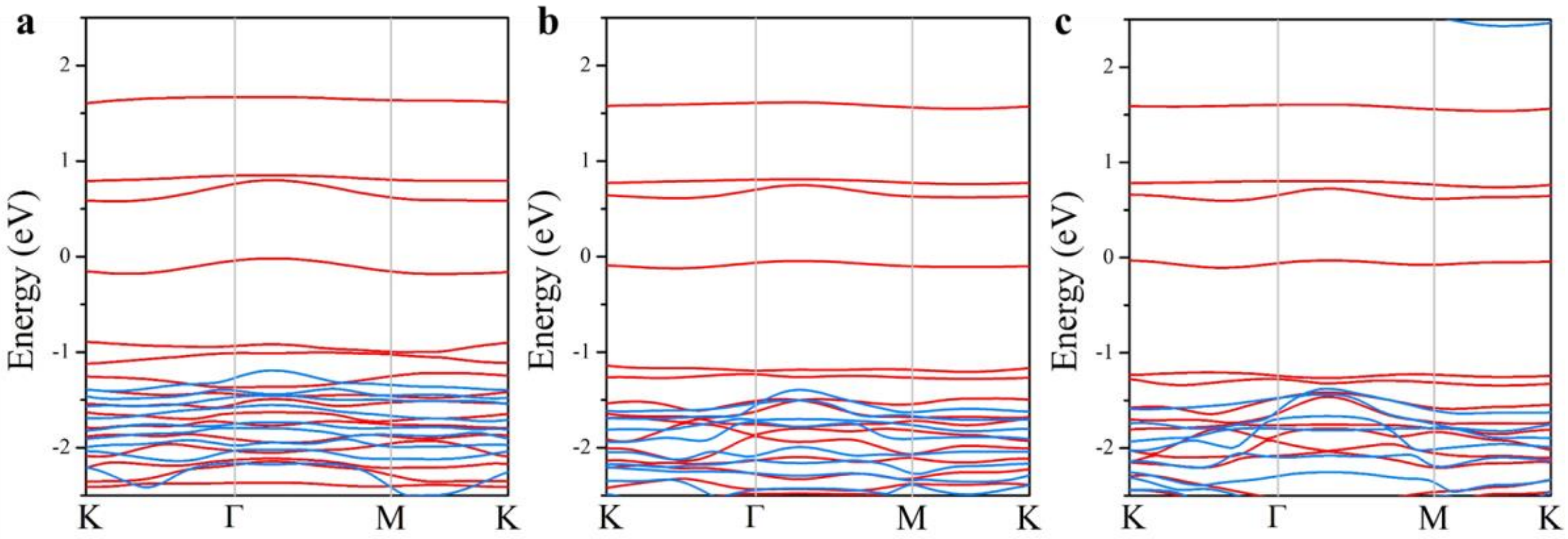

**Figure S7**. Band structures for (a) $(CrBr_3)_2NH_4$, (b) $(CrBr_3)_2Na$ and (c) $(CrBr_3)_2Li$. Red and blue lines represent spin up and spin down bands, respectively.

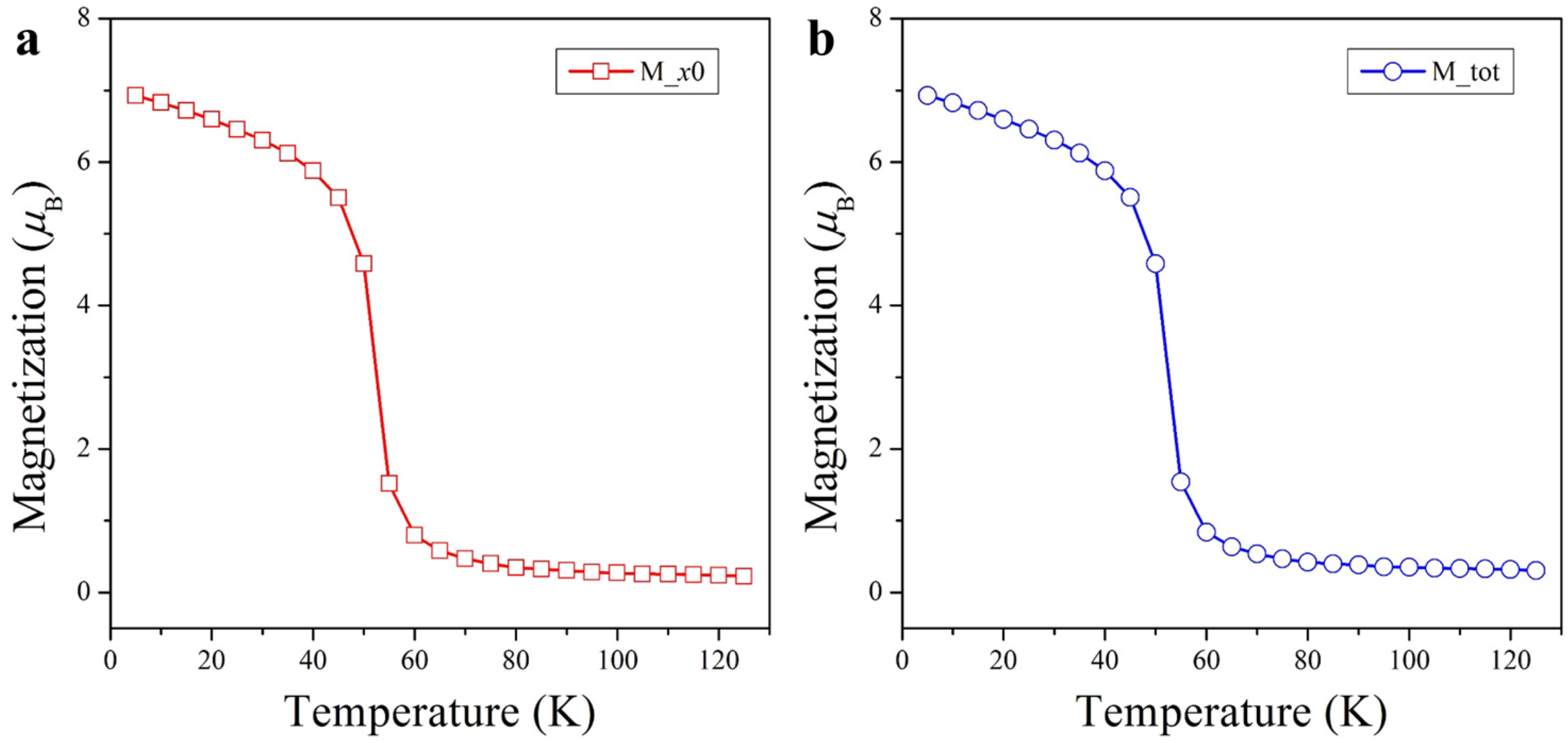

**Figure S8**. (a) Magnetic moment along easy-axis orientation and (b) total magnetic momnet per unitcell as a function of temperature from Monte-Carlo simulation for $(CrBr_3)_2Li$.

**Table S2**. The relative energies for different in-plane polarization orders of $(CrBr_3)_2NH_4$, $(CrBr_3)_2Na$ and $(CrBr_3)_2Li$. The AFE, FIE1 and FIE2 configurations are constructed in 2 × 1 supercells, and the 120° AFE and FIE3 configurations are constructed in √3 × √3 supercells. The FIE1 and FIE2 initial structures of these systems are always automatically converge into the corresponding FE and AFE structures. The FIE3 structure of $(CrBr_3)_2NH_4$ did not converge in the ionic relaxation process and its energy is much higher than the 120° AFE structure.

| Order | Relative energy per Cr (meV) | | |
|---|---|---|---|
| | $(CrBr_3)_2NH_4$ | $(CrBr_3)_2Na$ | $(CrBr_3)_2Li$ |
| FE | 40.8 | 7.9 | 0 |
| AFE | 42.1 | 17.1 | 21.3 |
| 120° AFE | 0 | 0 | 9.0 |
| FIE1 | -- | -- | -- |
| FIE2 | -- | -- | -- |
| FIE3 | -- | 17.2 | 22.3 |

## Section III: Magnetoelectic coupling.

The magnetoelectric (ME) coupling can be classified into two types: linear ME coupling and non-linear ME coupling. In the linear ME coupling case, the induced magnetization (**M**) is proportional to the applied electric field (**E**), or the induced electric polarization (**P**) is proportional to the applied magnetic field (**H**). The methods for computing the linear ME coupling tensor $\alpha$ were proposed in the literature [1,2].

In the non-linear ME coupling case, an external electric field larger than the coercive field causes the direction change of the electric polarization and a simultaneous flop of the magnetization. For example, the 71° or 109° switch of the electric polarization of $BiFeO_3$ thin-films leads to the flop of the magnetic easy-plane [3]. We note that our current system $(CrBr_3)_2Li$ also displays the non-linear ME coupling, i.e., the 60° or 120° switch of the electric polarization leads to the flop of the magnetic easy-axis. In the non-linear ME coupling case, although the change of magnetization is not proportional to the applied electric field in a strict manner, one can nevertheless define a ME coefficient to characterize the magnitude of the ME coupling: $\alpha \approx \mu_0|\Delta \boldsymbol{M}|/E_c$, where $\Delta \boldsymbol{M}$ is the change of magnetization before and after changing the direction of the electric polarization, and $E_c$ is the coercive electric field.

In the following, we will study the ME coupling in $(CrBr_3)_2Li$ with a method that was adopted to estimate the ME coupling coefficient of the Fe/$BaTiO_3$ system [4]. To estimate ME coefficient $\alpha$, one need to obtain $\Delta\boldsymbol{M}$ and the coercive electric field $E_c$. Without loss of generality, we now focus on the case of 120 ° switch of the electric polarization. With an approximate thickness (6 Å) of $(CrBr_3)_2Li$ layer, $\mu_0|\Delta\boldsymbol{M}| \approx 0.38$ Guass (G) if the magnetization is rotated by 60° or 0.66 G if the magnetization is rotated by 120°. The reason why there are two possible magnetization orientation is that the magnetization can be parallel or anti-parallel to the easy-axis when no external magnetic field is applied according to thermodynamics. The estimation of the coercive electric field $E_c$ is much more difficult since usually a real ferroelectric switching process is not homogenous, instead it involves domain-wall motion and/or defects [5]. Fortunately, it is well-known that the coercive electric field $E_c$ is related to the energy barrier ($U$) of the ferroelectric switching and the electric polarization ($P$) of the ferroelectric state: a smaller energy barrier and a larger electric polarization lead to a smaller coercive electric field $E_c$. Figure S9 shows the energy barrier for the 120 ° ferroelectric switching in $(CrBr_3)_2Li$ is 15 meV/unit-cell. We also check to see that the favored orientation of magnetization, i.e. the easy-axis, is indeed always parallel to the ferroelectric polarizations [see Fig. S10]. Considering the fact that $(CrBr_3)_2Li$ has a similar switching barrier and polarization ($U$ = 15 meV/unit-cell and $P$ = 15 $\mu$C/cm$^2$) as those ($U$ = 18 meV/unit-cell and $P$ = 28 $\mu$C/cm$^2$ from our test calculations) of bulk $BaTiO_3$, here we roughly take the experimental value of $E_c$ for $BaTiO_3$ (~10 kV/cm, see [6]) as the $E_c$ of $(CrBr_3)_2Li$. Then the ME coefficient $\alpha$ is estimated to be ~0.038 or ~0.066 G·cm/V. These values are comparable to that of Fe/$BaTiO_3$ (~0.01 G·cm/V from DFT calculations) [7] and $BiFeO_3$/$CoFe_2O_4$ (~0.01 G·cm/V from experiment) [8] multiferroic interfaces.

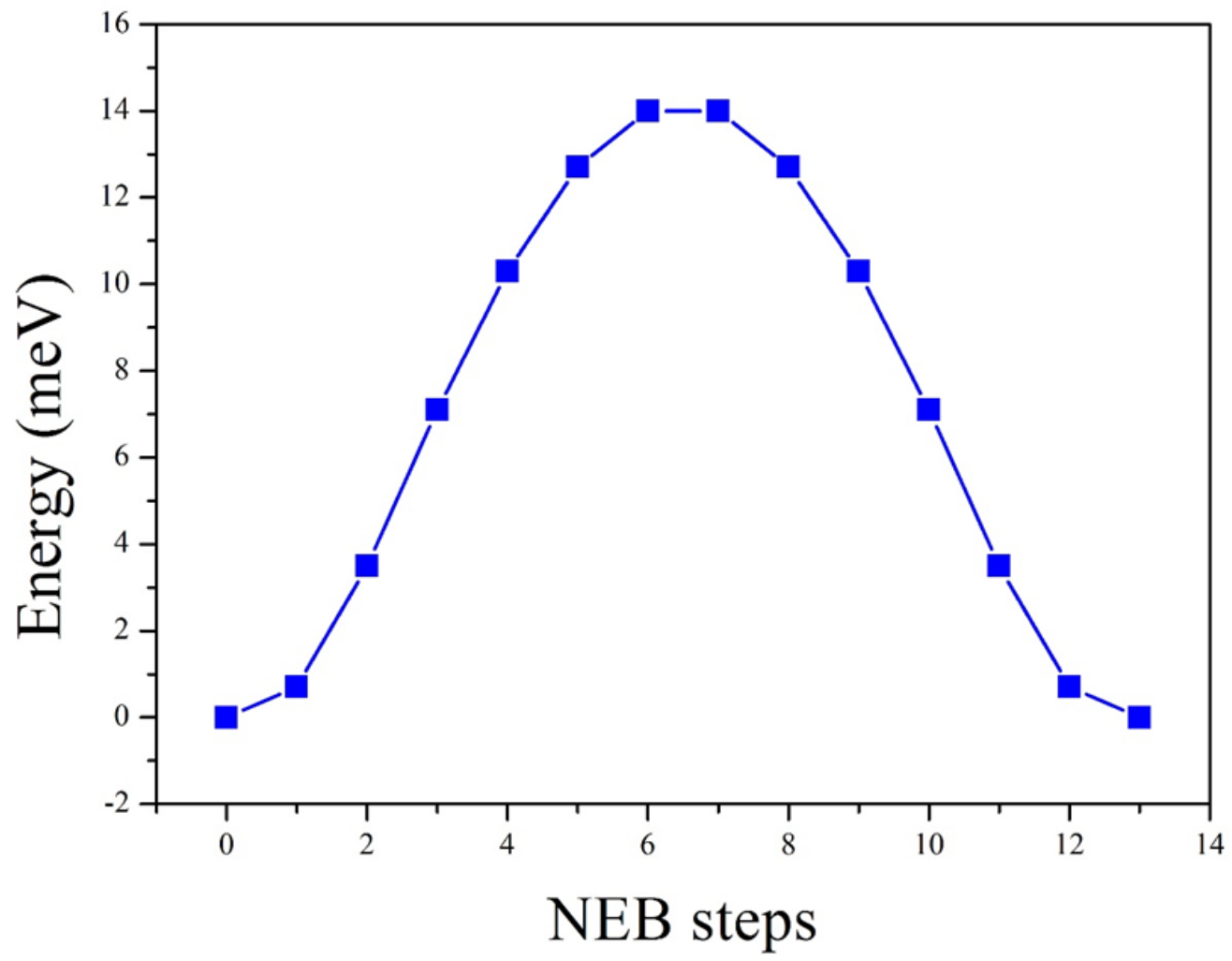

**Figure S9**. Energy (per atom) profiles for 120° ferroelectric switching in $(CrBr_3)_2Li$ from nudged-elastic-band (NEB) computation. $0^{th}$ and $13^{th}$ steps represent the FE structures with 0° and 120° in-plane polarizations, respectively. All the states during the NEB proceeding are insulating.

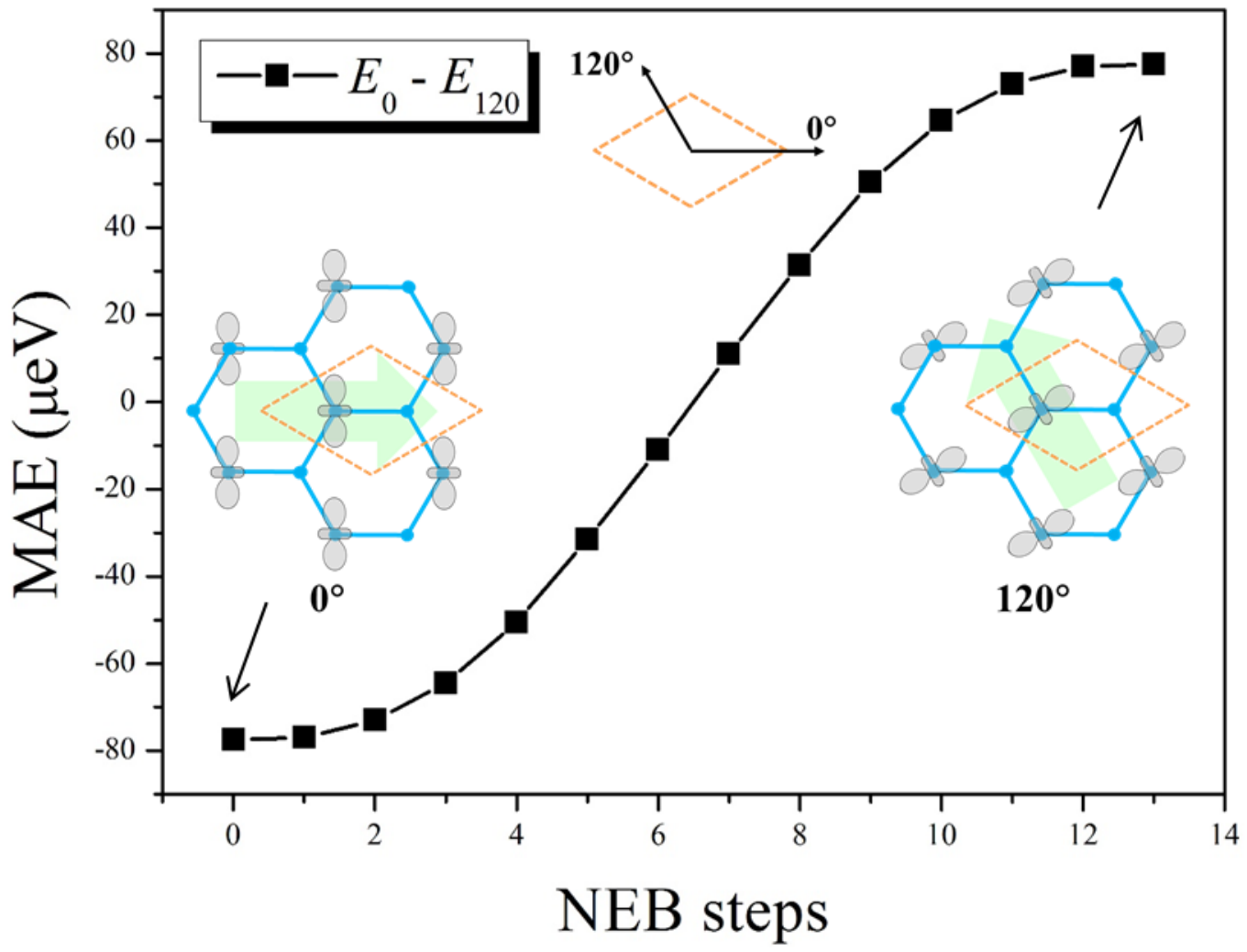

**Figure S10.** Variation of energy difference between 0° and 120° magnetizations during the 120° FE switching process. $0^{th}$ and $13^{th}$ steps represent the FE structures with 0° and 120° in-plane polarizations, respectively.

## Section IV: Substrate effect.

In practical applications, a proper substrate is always required to support a 2D material. Thus for experimental and practical interests, it is necessary to examine whether the inherent properties of a 2D material could be affected by the substrate. Here we choose two non-magnetic non-polar insulators, namely the β-tridymite $SiO_2$ and α-$InBr_3$ slabs to support the $(CrBr_3)_2Li$ monolayer. This is because $SiO_2$ is a widely used substrate in experiment to support 2D materials such as $CrI_3$ [9], and α-$InBr_3$ is an experimentally realized vdW layered material (ICSD 65198) and has very similar lattice structure as $(CrBr_3)_2Li$. The three-layer $SiO_2$(0001) and bilayer $InBr_3$ slabs are used to simulate the surfaces. The surfaces of $SiO_2$ are passivated by hydrogen. The lattice mismatches are 4.4% and 3.2% for the $(2\times2)SiO_2/(\sqrt{3}\times\sqrt{3})(CrBr_3)_2Li$ and $(\sqrt{3}\times\sqrt{3})InBr_3/(\sqrt{3}\times\sqrt{3})(CrBr_3)_2Li$ systems, respectively. The non-polar structure, i.e. the $D_{3d}$-$(CrBr_3)_2Li$ was used as the initial structure to perform atomic structure relaxations. After careful optimizations (including vdW correction), no strong chemical bonds are formed between $(CrBr_3)_2Li$ and the substrates (Fig. S11). The $(CrBr_3)_2Li$ layer always automatically converges into the ferroelectric $C_2$ phase, where the significant asymmetric JT distortions are observed. We also calculated the 120° antiferroelectric structure of $(\sqrt{3}\times\sqrt{3})(CrBr_3)_2Li$ on these substrates and find that it has a higher energy than the optimized FE structures. The relative energies of 120° antiferroelectric structures are 9 and 8 meV per Cr for $(\sqrt{3}\times\sqrt{3})(CrBr_3)_2Li$ on $SiO_2$ and $InBr_3$ substrates, respectively. These values are nearly same as that of isolated $(CrBr_3)_2Li$ monolayer (9 meV per Cr). Next we examined the magnetic and electronic properties of these systems. The ferromagnetic states are lower in energy than the antiferromagnetic states. The exchange energies ($E_{AFM} - E_{FM}$) are 13 and 18 meV per Cr for $(CrBr_3)_2Li/SiO_2$ and $(CrBr_3)_2Li/InBr_3$, respectively, which are also close to the value of isolated $(CrBr_3)_2Li$ monolayer (15 meV per Cr). The calculated band structures show that these systems are semiconducting. The Cr-$dz^2$ and Cr-$dx^2$-$y^2$ bands are located within the large energy gap of the substrates. Thus, we can expect that, the asymmetry JT distortions are robust against the substrates. The ferroelectricity, ferromagnetism and semiconducting features of the $(CrBr_3)_2Li$ monolayer can persist on a suitable substrate.

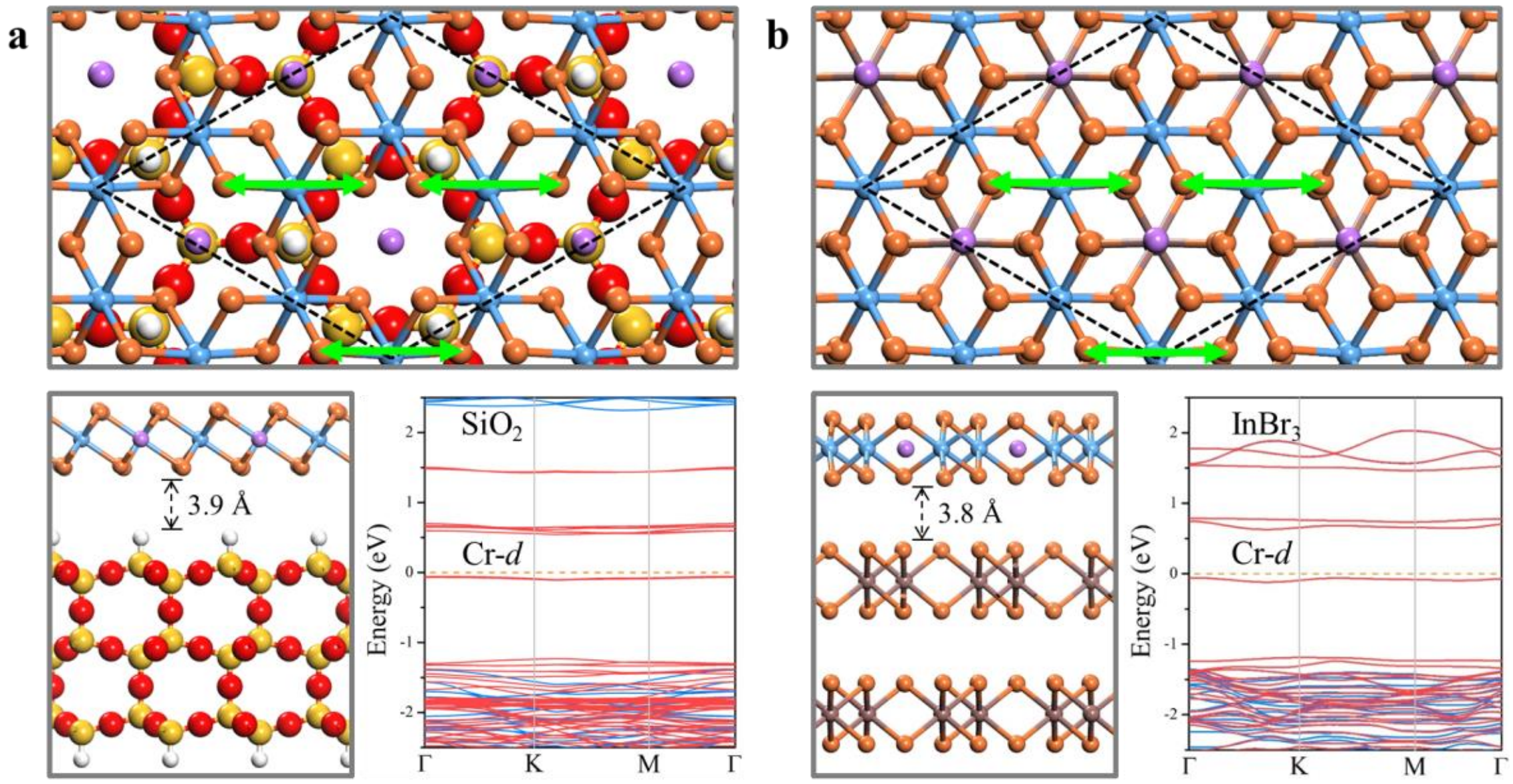

**Figure S11.** Top and side view of the optimized atomic structures and the band structures for (a) β-tridymite $SiO_2/(CrBr_3)_2Li$ and (b) α-$InBr_3/(CrBr_3)_2Li$ systems. Black-dot rhombuses represent the $\sqrt{3}\times\sqrt{3}$ unit cell of $(CrBr_3)_2Li$. Green double-headed arrows represent the asymmetric JT distortions, in which the Cr-Br bonds are longer than the others. Red and blue lines in the band structures represent spin-up and spin-down bands, respectively.

## Section V: Multiferroicity in other related 2D systems.

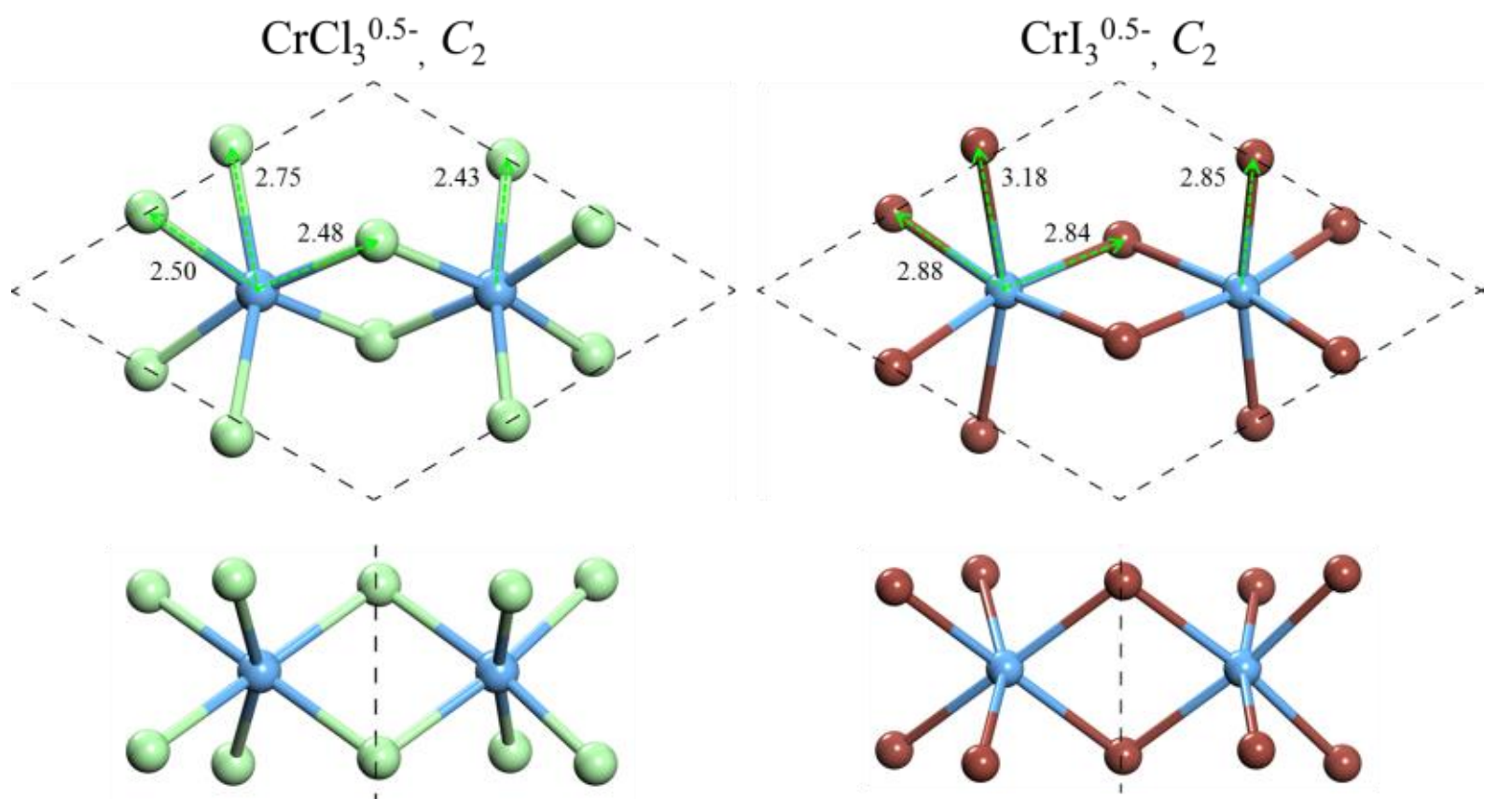

**Figure S12**. Optimized FE structures of $CrCl_3^{0.5-}$ and $CrI_3^{0.5-}$. The Cr-Cl and Cr-I bond lengths are in angstrom. The equilibrium 2D lattice constants are 6.63 and 7.64 Å, respectively.